\documentclass[9pt,sigconf,nonacm]{acmart}
\AtBeginDocument{%
  }

\usepackage{amsthm,amsmath}
\usepackage{algorithm,algorithmicx,algpseudocode}

\usepackage{graphicx}

\usepackage{booktabs}
\usepackage{multicol,multirow}
\usepackage{caption}
\usepackage{array}

\usepackage{pifont}

\usepackage{enumitem}
\setlist[itemize]{leftmargin=10pt}
\renewcommand{\arraystretch}{1}

\newcommand{\change}[1]{\textcolor{black}{#1}}

\begin{document}
\title{\underline{S}imultaneous \underline{Mu}lti-die \underline{F}loorplanning and \underline{T}echnology \underline{A}ssignment}

\author{Cristhian Roman-Vicharra$^1$ \quad Prianka Sengupta$^1$ \quad Runzhi Wang$^1$ \quad Yiran Chen$^2$ \quad Jiang Hu$^1$}
\affiliation{%
  \institution{$^1$Texas A\&M University, $^2$Duke University}
}
\email{{cristhianroman,prianka.sengupta,runzhi354,jianghu}@tamu.edu, yiran.chen@duke.edu}






\renewcommand{\shortauthors}{Roman-Vicharra et al.}

\begin{abstract}
In heterogeneous integration, different dies may employ distinct technologies, making floorplanning across multiple dies inherently coupled with technology assignment.
By assuming a fixed technology, almost all prior floorplanning studies were developed without addressing the challenge of technology assignment.
This work presents the first systematic study of \mbox{multi-die} floorplanning that treats technology choice as a variable.
To address the challenge of variable block areas, we incorporate a recent machine learning technique for rapid PPA estimation.
Our methods jointly optimize area, wirelength, performance, power, and cost, thereby highlighting the importance of technology assignment. Experimental evaluations, validated with a commercial tool for both 2.5D and 3D ICs, demonstrate that our systematic optimizations significantly outperform a greedy approach.
\end{abstract}


\keywords{Heterogeneous Integration, PPAC, Floorplanning, Technology Assignment}

\makeatletter
\def\@ACM@checkaffil{}
\makeatother
\maketitle

\vspace{-3mm}
\section{Introduction}
\label{sec:introduction}

Heterogeneous integration (2.5D chiplets and 3D ICs) enables opportunities for building complex IC designs into a single chip with applications in high-performance computing, 5G, and AI fields.
Integration involves multiple dies often from different manufacturing technologies.
Compared to monolithic ICs, heterogeneous integration also allows the reuse of existing IPs (Intellectual Properties), and achieves enhanced functionality, compact area, and design flexibility.
Multi-die floorplanning for 2.5D/3D IC systems has been extensively studied~\cite{ho2013btreeSA,lee2023multiDieFloorplan,osmolovskyi2018diePlacement,ma2021thermalFloorplan,hsu2022warpageFloorplan,duan2024RLPlanner,amin2024RLFloorplan3D,chen2024floorplet,hwang2011tsvFloorplan,want2013tsvCoPlace}. These studies primarily focus on optimizing metrics such as chip area, interconnect (wirelength and TSVs), timing, thermal, or warpage.
Typical approaches include simulated annealing, branch-and-bound, and reinforcement learning.


Prior work explored performance metrics without considering the significant challenge of technology assignment---a critical industrial need.
In heterogeneous integration, floorplanning is tightly coupled with technology assignment because different dies may use different technologies.
Conventional floorplanning assumes predefined block technologies and dimensions, whereas heterogeneous systems treat these attributes as design variables.
Moreover, pre-assigning a block to a technology is problematic because it constrains the system without knowledge of the implications for interconnect, timing, and other metrics.
Performing floorplanning and technology assignment simultaneously greatly expands the solution space, introducing additional challenges in search efficiency and optimization.
Indeed, multi-die and multi-technology floorplanning is recognized as a major challenge in heterogeneous integration~\cite{chang2024multidieChallenges,bei2025chipletChallenge}.
Despite extensive prior work on 2.5D and 3D floorplanning, the joint consideration of technology assignment remains almost entirely unaddressed.


We propose a straightforward approach to solve the Simultaneous Multi-die Floorplanning and Technology Assignment (SMUFTA) problem.
Our SMUFTA employs with simulated annealing, Bayesian optimization, and reinforcement learning to simultaneously optimize multiple objectives, including performance (total negative slack), power, area, die cost, and total wirelength (intra-die and inter-die connections).
Furthermore, SMUFTA integrates recent ML techniques~\cite{sengupta2022ppa} for technology-specific PPA estimation of circuit blocks, and also accommodates both soft (synthesizable HDL) and hard (layout) IPs.
The method is validated through full-system integration in commercial tools, using a silicon interposer for 2.5D ICs and Through-Silicon Vias (TSVs) for 3D ICs.


The SMUFTA problem for 2.5D/3D heterogeneous systems is a complex engineering challenge in IC design.
Due to its broad scope, it is impractical to address all associated issues in a single study.
Thus, this work is specifically focused on PPAC (Performance, Power, Area, and Cost) optimization.
Other critical aspects, such as thermal effects, mechanical stress, and reliability, will be addressed in our future research.

\noindent
{\bf Our primary contribution.}
Given that no prior research has addressed the SMUFTA problem, the primary value of this work is not to outperform existing methods.
Our main contribution is the first systematic study of this critical problem, which has received little research attention despite its growing importance with the adoption of chiplets/3D ICs.
Other contributions include the following:
\vspace{-1mm}
\begin{itemize}

\item \change{We present a straightforward solution to the chicken-and-egg dilemma~\cite{chang2024multidieChallenges} in the simultaneous multi-die floorplanning and technology assignment.}


\item \change{This work studies three optimization techniques including a state-of-the-art reinforcement learning approach.}

\item To ensure credible assessments, techniques are evaluated using a commercial tool across various experimental settings.


\item \change{Results demonstrate that a simple greedy approach is insufficient, showing the need for sophisticated methods (such as reinforcement learning) to achieve competitive PPAC results.}

\end{itemize}

\vspace{-3mm}
\section{Previous Works on Multi-die Floorplanning}
\label{sec:previousWorks}

The challenges in heterogeneous integration and multi-die/multi-technology floorplanning are significant as discussed in~\cite{chang2024multidieChallenges}. The Simulated Annealing (SA) algorithm is a common optimization approach for floorplanning~\cite{ho2013btreeSA,lee2023multiDieFloorplan,hsu2022warpageFloorplan,ma2021thermalFloorplan}.
The study in~\cite{ho2013btreeSA} minimized wirelength and area for interposer-based chiplet floorplanning but drawbacks are that the number of chiplets \change{is} limited and long runtime.
The methodology in~\cite{lee2023multiDieFloorplan} also optimized wirelength and area while including multi-die interconnect bridges, but lacking internal die floorplanning.
The work of~\cite{hsu2022warpageFloorplan} proposed a heterogeneous floorplanning method to mitigate warpage during the packaging process.
The approach in~\cite{ma2021thermalFloorplan} introduced thermal-aware chiplet floorplanning to minimize operating temperature and wirelength using thermal simulators. However, it relies on space insertion between chiplets potentially increasing chip.

Reinforcement Learning (RL) is adopted by modelling the floorplanning as a Markov Decision Process (MDP)~\cite{xu2022goodfloorplan,amin2024RLFloorplan3D,duan2024RLPlanner}.
GoodFloorplan~\cite{xu2022goodfloorplan} uses an RL framework with graph convolutional networks, outperforming SA-based methods in area and wirelength.
Another RL-based framework~\cite{amin2024RLFloorplan3D} uses a decision transformer to optimize wirelength, congestion, and heat, allowing to prompt desired objective values.
Similarly, RLPlanner~\cite{duan2024RLPlanner}, inspired in~\cite{ma2021thermalFloorplan}, optimizes wirelength and temperature by integrating a fast physics-informed thermal evaluator.

Beyond SA and RL, methods using Mathematical Programming (MP) have been proposed. Floorplet~\cite{chen2024floorplet} is a performance-aware approach that optimizes wirelength, warpage, and packaging cost by incorporating yield, warpage, and bump stress models. Floorplet utilizes the yield model introduced in ~\cite{feng2022costModel,cunningham1990costModel}.
The work in~\cite{zhuang2022multipackage} is used for multi-package co-design integration to optimize wirelength, warpage, bump, and interconnection cost, while maintaining non-overlapping and bump constraints. but it requires large runtime.


The planning and placement of TSVs significantly impact performance metrics and total wirelength in 3D-ICs.
Prior work includes a TSV-aware floorplanning method~\cite{hwang2011tsvFloorplan} that primarily reduces wirelength using SA-based TSV planning and flow network-based reassignment.
The study in~\cite{want2013tsvCoPlace} proposes 3D floorplanning with TSV co-placement to optimize wirelength and TSV count, also incorporating a TSV reassignment step.
Despite extensive efforts in 2.5D/3D floorplanning, almost no prior work has simultaneously considered technology assignment.
Thus, the multi-die multi-technology assignment challenge~\cite{chang2024multidieChallenges} remains largely unsolved.
\vspace{-2mm}
\section{Background}
\label{sec:background}
\noindent
{\bf PPA (Total Negative Slack, Power, Area) Estimation.}
The PPA metrics for a circuit block are estimated using XGBoost-based ML models introduced in~\cite{sengupta2022ppa}.
These models are technology-specific, with one model for each technology.
The models are built using post-placement analysis data (obtained using vectorless analysis) and use input features such as HDL-based and synthesis parameters.
We extended the work of~\cite{sengupta2022ppa} by incorporating area estimation.

\noindent
{\bf PPO (Proximal Policy Optimization) Algorithm.}
PPO~\cite{schulman2017PPOclip} is a popular RL algorithm and is adopted in OpenAI's ChatGPT training.
The advantages of PPO are its simplicity, stability and efficiency, while often being computationally less expensive than other RL algorithms.

\noindent
{\bf B*-tree Representation.} 
The B*-tree~\cite{chang2000btree} is an ordered binary-tree representation for non-slicing floorplans, offering advantages over other representations~\cite{wong1986polishExpression,murata2003sequencePair,nakatake1996boundSlicing,guo1999oTree,hong2000cornerList}.
It provides a 1-to-1 transformation between a floorplan and its B*-tree in linear time. 


\vspace{-2mm}
\section{Problem Formulation}
\label{sec:formulation}
\noindent 
{\bf SMUFTA 2.5D} (\underline{S}imultaneous \underline{Mu}lti-die \underline{F}loorplanning and \underline{T}echnology \underline{A}ssignment):
Given a circuit system composed of a set of interconnected blocks $\mathcal{B}$ described by synthesizable HDL code, aspect ratio options for each block, an ordered set of technologies 
$g_1, g_2, ...,g_k$,
and a number of silicon dies $\mathcal{D}$ each having a distinct technology node, SMUFTA assigns blocks to the given dies and determines their locations on the dies along with their aspect ratios to
\begin{align}
 \text{minimize}~~ & f = \omega \cdot W + \beta \cdot P+\gamma \cdot \sum_{d_i \in \mathcal{D}} C(d_i) + \tau \cdot T \label{eq:objective}\\
 \text{subject to}~~ & N_{i,j} \le N_{max},~~ d_i, d_j \in \mathcal{D}, i\ne j  \\
  & A_{min} \le A(d_i) \le A_{max},~~ d_i \in \mathcal{D},  \label{eq:area_constraint}
\end{align}
where $W$ is the total Half-Perimeter Wire-Length (HPWL), $P$ is the total dynamic power of all blocks, $C(d_i)$ is the cost of die $d_i$ based on its yield model, $T$ is the absolute sum of the total negative slack (TNS) across all blocks, $N_{i,j}$ is the number of nets \change{connecting} dies $d_i$ and $d_j$, $A(d_i)$ is the area of die $d_i$, and $\omega$, $\beta$, $\gamma$, $\tau$, $N_{max}$, $A_{min}$ and $A_{max}$ are constant parameters.

\begin{figure}[!ht] 
\centering
\includegraphics[width=.8\linewidth]{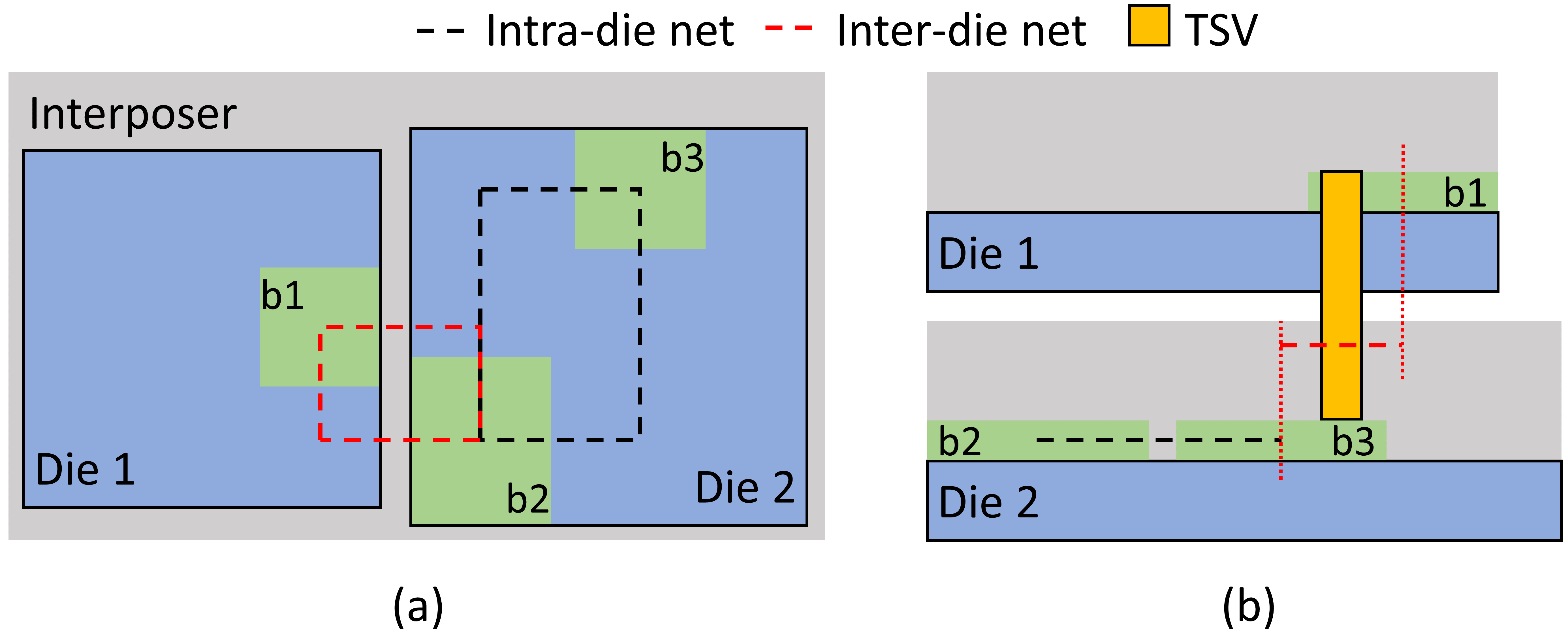}
\vspace{-3mm}
\caption{(a) Top view of a inter-die net in 2.5D chiplet and (b) side view of a TSV net in 3D formulation.}
\vspace{-3mm}
\label{fig:wire-formulation}
\end{figure}

The HPWL $W$ assumes that the pins of a \change{rectangular} block are at its center, as in many previous works on floorplanning, and it includes both inter-die (interposer) and intra-die wirelength. For example, Figure~\ref{fig:wire-formulation}a shows an inter-die net between two blocks in different dies.
Area is considered in the constraints, not in the objective function, becaue it is correlated with $W$, $C$ and $P$.
This formulation is targeted to 2.5D chiplet integration using silicon interposer. However, it is also applicable to multi-die InFO packaging~\cite{feng2022costModel}.

\noindent
{\bf SMUFTA 3D:} The wirelength of inter-die nets in a 3D formulation differs because the dies are vertically aligned (Figure~\ref{fig:wire-formulation}b). The vertical connection between inter-die nets is addressed using TSVs.
The HPWL $W$ includes both the intra-die and TSV-connected wirelength as if the blocks are placed in the same horizontal plane.
The assumption is that the lower-left coordinates of each die are located at the origin coordinate.

\noindent
{\bf SMUFTA with hard IPs.} A special case of SMUFTA occurs when some circuit blocks are hard IPs, and therefore their technologies and aspect ratios are fixed throughout the optimization.
Hard IPs implies that a given block requires a specific technology, limiting its options for die assignment based on the fixed technology.

\vspace{-2mm}
\section{The Proposed Method}
\label{sec:method}
\begin{figure}[ht] 
\centering
\includegraphics[width=.9\linewidth]{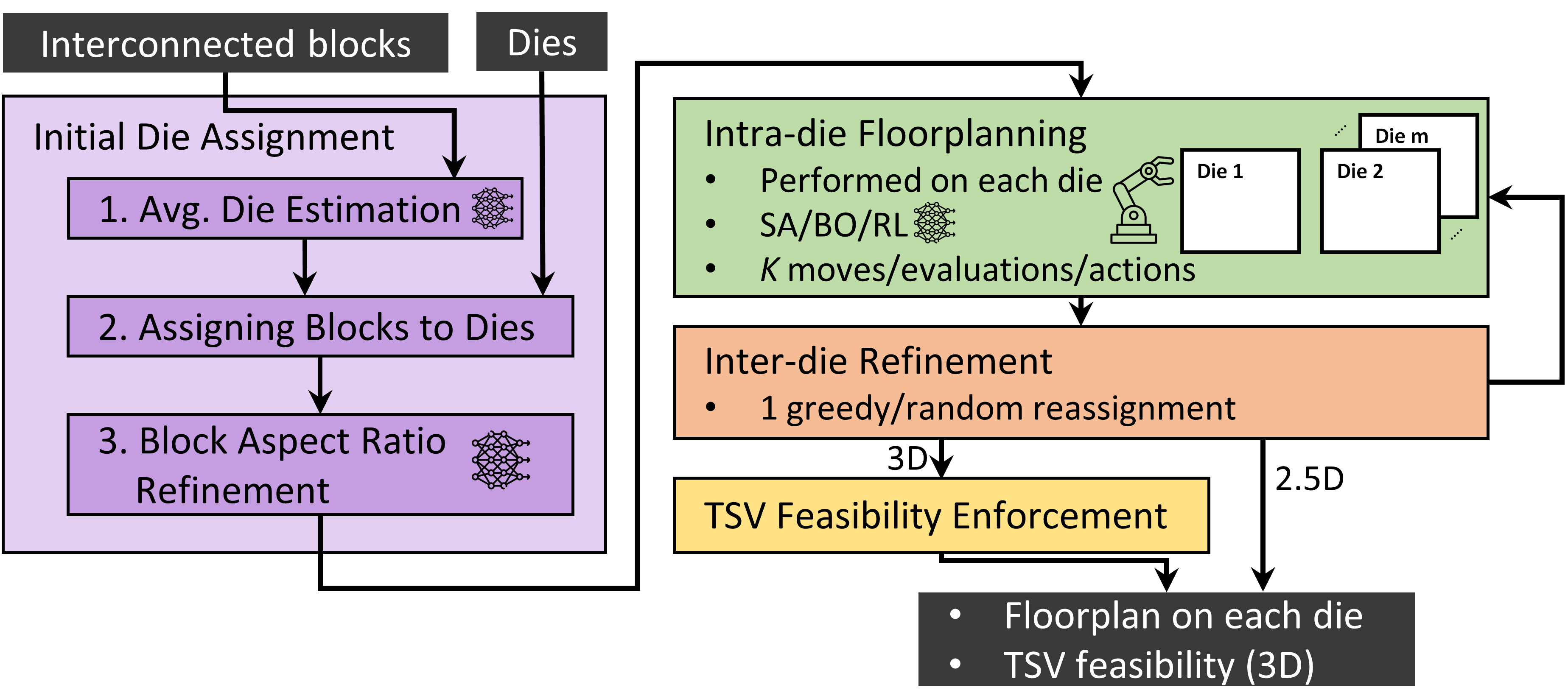}
\vspace{-3mm}
\caption{An overview of the proposed SMUFTA methodology.}
\vspace{-5mm}
\label{fig:workflow}
\end{figure}

An overview of the proposed SMUFTA methodology is provided in Figure~\ref{fig:workflow}. It consists of initial die assignment, intra-die floorplanning, inter-die refinement, and TSV feasibility enforcement for 3D-ICs.

\vspace{-2mm}
\subsection{Area, Cost and Wirelength Models}
\label{sec:models}
\subsubsection{Die Area}
\label{sec:area_model}
The area $A(d_i)$ of a die $d_i \in \mathcal{D}$ is the area of the minimum bounding box enclosing its block floorplan margin area.

\subsubsection{Die Cost Model}
\label{sec:cost_model}
The manufacturing yield of a single die, $Y(d)$, is crucial for determining die cost.
In~\cite{feng2022costModel,cunningham1990costModel}, the yield model is defined as
\begin{equation}
  Y(d) = \left(1+ \frac{\delta\cdot A(d)}{\alpha} \right)^{-\alpha},
\end{equation}
where $A(d)$ is the area of die $d$, $\delta$ is the defect density, and $\alpha$ is a model parameter.
For example, $\delta$ is $0.09$~cm$^{-2}$ and $\alpha$ is $10$ for $7$nm technology in~\cite{feng2022costModel}.
We only consider the manufacturing cost per yield area, $C(d){=}\Phi/Y(d)$ as cost model, where $\Phi$ is a technology-dependent constant.
Packaging and interconnect costs are not explicitly included, but are partially addressed as packaging is proportional to the total die cost, and interconnect is related to inter-die/TSV wirelength.

\subsubsection{Wirelength Model}
A \change{multi-terminal} net $e$ is a subset of blocks $e\subseteq \mathcal{B}$.
The HPWL of net $e$ is defined as
\begin{equation}
  W_e = \max_{b_i \in e}{x_i}-\min_{b_i \in e}{x_i} + \max_{b_i \in e}{y_i}-\min_{b_i \in e}{y_i},
\end{equation}
where $x_i$ ($y_i$) is the horizontal (vertical) coordinates of the center of block $b_i$.
The HPWL $W$ includes the intra-die and inter-die wirelength in Section~\ref{sec:formulation}.

\vspace{-2mm}
\subsection{Phase I: Initial Die Assignment}
\label{sec:initial_assig}
This stage aims to evenly assign the $n$ given blocks in $\mathcal{B}$ to the $m$ dies in $\mathcal{D}$ across $k$ technologies ($k \le m$), such that the subsequent floorplanning may converge faster than a random initial solution.
The area of a block $b_i$ is denoted as $A(b_i, g_i, \rho_i)$, where $g_i$ is the assigned technology and $\rho_i$ is its aspect ratio. This area is estimated using the ML model described in Section~\ref{sec:background}.
Note that the aspect ratio $\rho_i$ is a layout tool parameter used as an input feature to the ML models.

\noindent{\bf Step 1: Average die area estimation.}
Assuming the $k$ technologies $g_1, g_2, ..., g_k$ are ordered from oldest (largest) to newest (smallest).
This step estimates the average die area $z$ for all dies in $\mathcal{D}$ to accommodate all blocks in $\mathcal{B}$. 
To do this, we temporarily assign all $n$ blocks to the oldest technology $g_1$ with an aspect ratio of $1$. We then normalize the area $z$ of every die to technology $g_1$ by using a scaling factor $s_i$ between a $g_i$ and $g_1$
We assume area scales quadratically with technology feature size (e.g., $s_2=4$ for $g_2=7$nm and $g_1=14$nm).
To ensure that all dies have approximately the same $z$ area in their respective technologies, we enforce the linear equation
\[
\sum_{i=1}^m s_i \cdot z = \sum_{i=1}^n A(b_i, g_1, 1).
\]
While the scaling factors $s_i$ are ideal, they are only used in this step to obtain a preliminary area estimate.

\noindent{\bf Step 2: Assigning blocks to dies.}
All blocks in $\mathcal{B}$ are sorted in non-increasing order of their area $A(b_i, g_1, 1)$ in the oldest technology $g_1$ with an aspect ratio of 1.
Separately, all dies in $\mathcal{D}$ are sorted in increasing order of their technology node (newest to oldest).
Following the ordered sets, blocks are assigned one-by-one to the first die until the total block area approximately reaches $z$. This process is repeated for the remaining blocks and subsequent dies until all blocks are assigned. Note that once block $b_i$ is assigned to a die with technology $g_j$, its area becomes $A(b_i, g_j, 1)$.

\noindent{\bf Step 3: Block aspect ratio refinement.}
After all blocks are assigned to dies, we determine each block's aspect ratio ($\rho_i$) to minimize
\begin{equation}
  f_{pt} = \beta \cdot P + \tau \cdot T, \label{eq:part_objective}
\end{equation}
which is a partial objective in Equation~\eqref{eq:objective}. $f_{pt}$ is computed for each block using the ML models (Section~\ref{sec:background}) across a set of aspect ratios ($0.8, 0.9, 1.0, 1.1, 1.2$).
Then, the ratio that minimizes $f_{pt}$ is selected as $\rho_i$ for block $b_i$.
Although floorplanning has not been performed yet, Phase~I provides a good initial solution concerning TNS and power.

Once Phase~I is completed, each block $b_i \in \mathcal{B}$ is assigned a die $d_i \in \mathcal{D}$, a technology $g_i$, and an initial aspect ratio $\rho_i$.
Note that these aspect ratios are subject to change in the subsequent Phases.

\vspace{-2mm}
\subsection{Phase II: Intra-Die Floorplanning}
\label{sec:intradie_floorplan}
The intra-die floorplanning consists of placing disjoint subsets of blocks $\mathcal{B}_i$ into each die $d_i \in \mathcal{D}$.
We utilize the B*-tree representation~\cite{chang2000btree} because its efficiency and ability to handle non-slicing floorplan (Section~\ref{sec:background}).
This optimization is performed for all dies using three approaches: Simulated Annealing (SA), Bayesian Optimization (BO), and Reinforcement Learning (RL), which execute a move, an evaluation, and an action within the B*-tree, respectively.
Phase~II aims to improve the floorplan solution within each die $d_i$.

\subsubsection{Simulated Annealing-based Floorplanning}
\label{sec:sa}
The initial solution is a randomly created B*-tree using the blocks $\mathcal{B}_i$. An SA move includes the following perturbations to the B*-tree representation:
\vspace{-1mm}
\begin{itemize}
  \item Swapping two nodes to primarily reduce HPWL $W$ and cost $C(d_i)$.
  \item Remove-and-insert, removing and inserting a node to a leaf, to diversify the floorplan solutions.
  \item Changing the aspect ratio of a block to mainly reduce TNS $T$, power $P$, and cost $C(d_i)$.
  \item Rotation of a block to reduce die area $A(d_i)$.
\end{itemize}
The objective function $f$ is the same as in Equation~\eqref{eq:objective} for a single die $d_i$.
The SA algorithm uses standard hyperparameters (initial temperature, cooling factor, and convergence stopping criteria).
This SA approach provides a first attempt on solving the SMUFTA problem.

\subsubsection{Bayesian Optimization-based Floorplanning}
\label{sec:bo}
We implement a Bayesian Optimization (BO) approach as an alternative to the SA local search. The elements of the BO-based optimization are:
\vspace{-1mm}
\begin{itemize}
  \item The search space: The set of possible B*-tree configurations, block aspect ratios, and block orientations.
  \item Probabilistic surrogate model (i.e. Gaussian Process) for global approximation of the objective function $f$ as in Equation~\eqref{eq:objective}.
  \item Acquisition function (i.e. Expected Improvement) to select the next B*-tree configuration, balancing exploration and exploitation.
\end{itemize}
The BO hyperparameters are the optimization budget and the number of initial samples.
This BO approach provides a sample-efficient method on addressing the SMUFTA problem.

\subsubsection{Reinforcement Learning-based Floorplanning}
\label{sec:rl}
\begin{figure}[ht]
\centering
\vspace{-2mm}
\includegraphics[width=.9\linewidth]{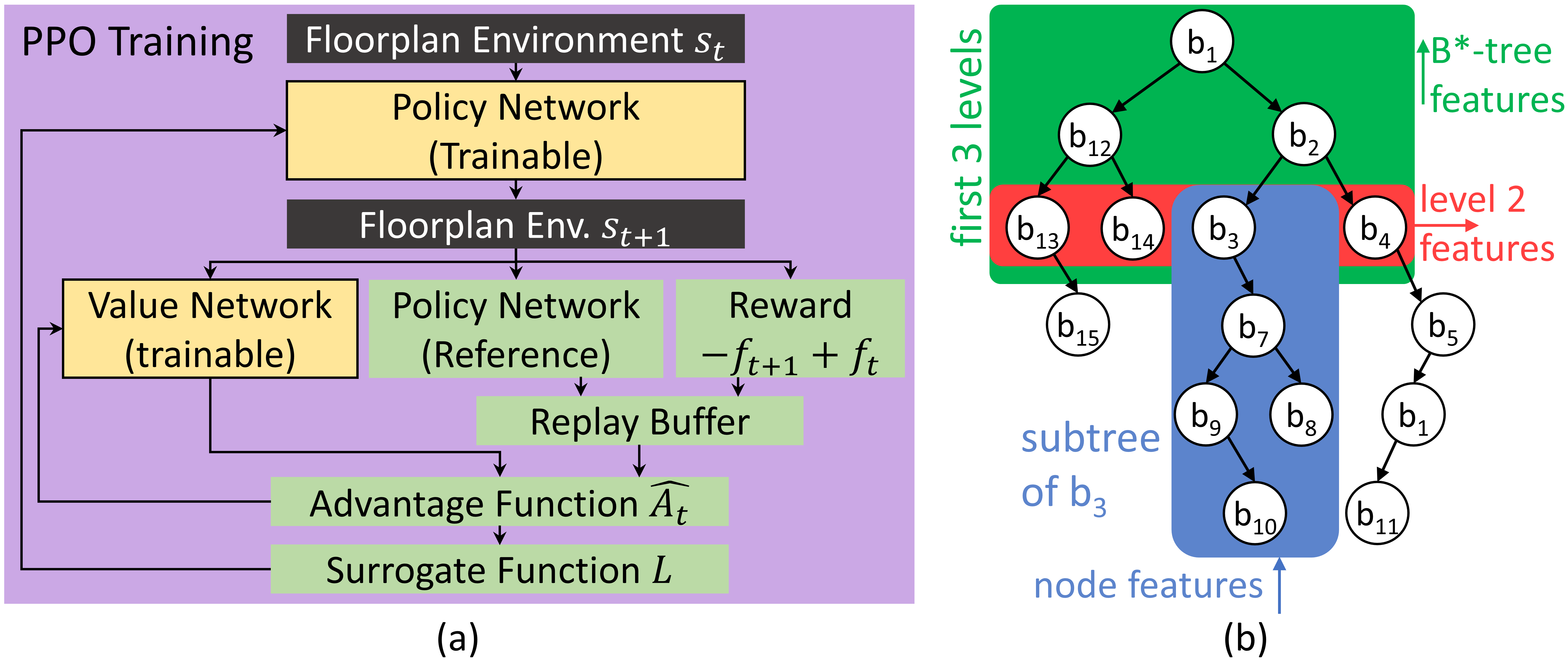}
\vspace{-3mm}
\caption{(a) Policy and Value Networks training for RL. (b) The B*-tree feature extraction.}
\vspace{-4mm}
\label{fig:networks}
\end{figure}

We adopt the PPO algorithm (Section~\ref{sec:background}) for our RL approach.
The key elements of the RL-based floorplanning environment are:
\vspace{-1mm}
\begin{itemize}
  \item State space $\mathcal{S}$: The set of possible B*-tree configurations of blocks $\mathcal{B}_i$ in die $d_i \in \mathcal{D}$. A state $s \in \mathcal{S}$ is the B*-tree of a floorplan solution.
  \item Action space $\mathcal{A}$: The discrete set of perturbations to a B*-tree as defined earlier in Section~\ref{sec:sa}.
  \item Reward function $R_a(s,s'){=}-\left(f_{t+1}-f_t\right)$: The negative difference in the objective $f$ when transitioning from state $s$ to $s'$ at episodic time $t$. The value of $f$ is evaluated using models in Section~\ref{sec:background}~and~\ref{sec:models}.
\end{itemize}
The policy and value functions are estimated using fully-connected (FC) neural networks trained throughout several episodes of intra-die floorplanning (Figure~\ref{fig:networks}a).
The policy network uses two FC layers with ReLU and a final FC layer with Softmax to obtain a probability distribution.
Similarly, the value function network uses three FC layers with ReLU and a single output neuron for a scalar value.
Both networks, colored yellow in Figure~\ref{fig:networks}a, use the same input features collected from a state $s$ in the floorplanning environment.
Then, we select B*-tree statistics (B*-tree features) as input features following:

\noindent{\bf Node features.}
These features consist of the height, the number of \change{cumulative} right children, the number of \change{cumulative} left children, the total number of nodes, and the HPWL of the blocks within the corresponding sub-tree. For example, in Figure~\ref{fig:networks}b, the node features of $b_3$ are a height of 3, 3 right children \change{($b_7, b_8, b_{10}$)}, 1 left child \change{($b_9$)}, 5 total nodes, and the HPWL value of the sub-tree (highlighted blue).

\noindent{\bf Level features.}
The level features are extracted by averaging the node features for all nodes at the same level in the B*-tree. For instance, the level-2 features in Figure~\ref{fig:networks}b are the averaged node features of blocks $b_{13}, b_{14}, b_3$, and $b_4$ (highlighted in red).

\noindent{\bf B*-tree features.}
These features are the concatenation of level features from the first $h$ levels forming a one-dimensional vector of size $5\cdot h$ in Figure~\ref{fig:networks}b (highlighted green when $h=3$).

Our RL approach performs intra-die floorplanning by having a dedicated agent for each die and its corresponding technology.
Each agent leverages technology-specific properties into the parameters $\theta$ and $\phi$ of its policy and value networks, respectively.
As a result, the RL approach is more efficient than SA's random moves or BO's expensive surrogate modeling.
The effectiveness of all proposed methods is significantly enhanced by leveraging the models for PPA estimation.

\vspace{-2mm}
\subsection{Phase III: Inter-Die Refinement}
\label{sec:refinement}
After every $K$ moves (SA), evaluations (BO), or actions (RL) in Phase~II, a global inter-die refinement is performed.
This step balances exploration and exploitation by randomly or greedily selecting a block $b$ from die $d_i$ for reassignment to die $d_j$. 
The reassignment is only accepted if the area constraint in Equation~\eqref{eq:area_constraint} is satisfied.
Otherwise, another reassignment is evaluated.
The block $b$ is then inserted into the B*-tree of $d_j$ with its position remaining near the boundary of $d_i$.

The inter-die refinement interleaves random and greedy selections.
The greedy selection promotes exploitation by choosing the block that achieves the maximum reduction in the partial objective $f_{pt}$ as Eq.~\eqref{eq:part_objective}.
Conversely, the random selection facilitates exploration.
The choice of the parameter $K$ (frequency of refinement) is a design choice as there is a sweet spot.
A large $K$ may saturate objective improvement, while a small $K$ may lead to unnecessary computation.
\vspace{-4mm}
\subsection{Phase IV: TSV Feasibility Enforcement}
\label{sec:tsv_placement}

In 3D ICs, \change{TSVs handle the connections between vertically stacked dies}. This Phase aims to facilitate feasibility by ensuring enough space for TSV nets and avoiding HPWL overhead during TSV placement. The process of allocating space follows:

\noindent{\bf TSV Area Demand.}
The required area, for TSVs connecting two blocks $b_i$ and $b_j$, is calculated as $\#net_{i,j} \cdot A_\text{TSV}+\mu$, where $\#net_{i,j}$ is the number of nets between $b_i$ and $b_j$, $A_\text{TSV}$ is the area of a single TSV, and $\mu$ is a constant margin.

\noindent{\bf Internal Empty Space.}
The available empty space $(1-\kappa_i) {\cdot} A(b_i) {\cdot} \gamma_{i,j}$ within block $b_i$ is computed based on its area $A(b_i)$, standard-cell density $\kappa_i$, and overlap ratio $\sigma_{i,j}$ (between $b_i$ and the TSV-net bounding box). The density $\kappa_i$ is a tool parameter used during placement.

\noindent{\bf Allocate White Space.}
We allocate TSVs within the internal empty space. If the demand exceeds the empty space, we insert white space on a side of $b_i$ contiguous to $b_j$. The inserted white space also is available for other TSV nets.

\begin{figure}[ht]
\centering
\vspace{-3mm}
\includegraphics[width=.7\linewidth]{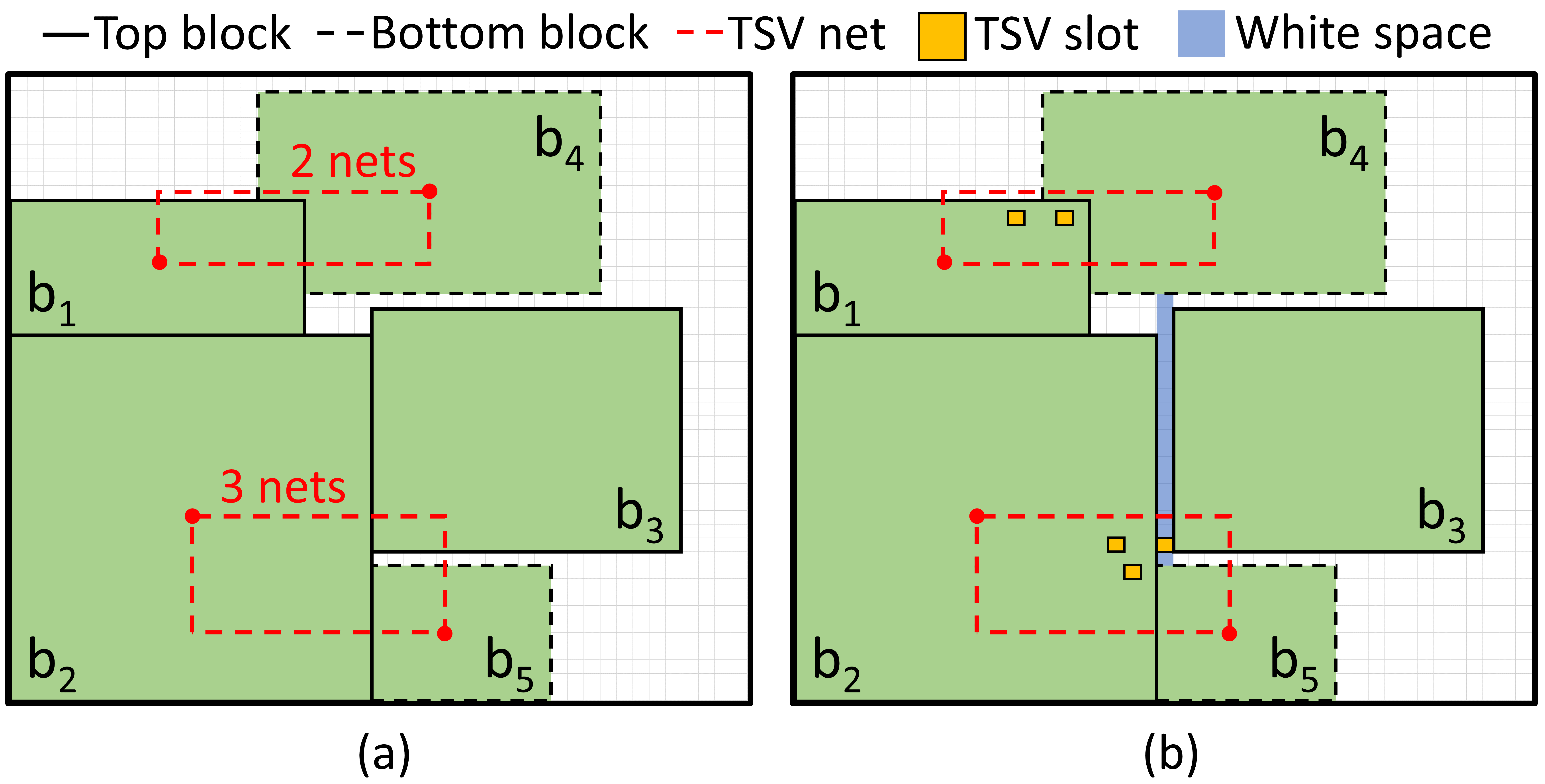}
\vspace{-3mm}
\caption{TSV Feasibility enforcement in 3D ICs. (a) top-down view of five TSV nets and (b) white space allocation.}
\vspace{-3mm}
\label{fig:tsv_feasibility}
\end{figure}

Figure~\ref{fig:tsv_feasibility} illustrates the bounding boxes for TSV nets. Between $b_1$ and $b_4$, if the empty space covers the demand, TSVs are placed within $b_1$. Conversely, between $b_2$ and $b_5$, if empty space cannot satisfy the demand, white space is inserted on the right side of $b_2$ \change{contiguous to $b_3$. Note that Figure~\ref{fig:tsv_feasibility} is only intended to represent the white space allocation and not a floorplan solution.}

\vspace{-2mm}
\section{Experiments}
\label{sec:experiments}

\subsection{Experiment Setup}
\begin{table}[H]
  \centering
  \vspace{-3mm}
  \caption{Designs for Optimization}
  \vspace{-3mm}
  \resizebox{0.5\linewidth}{!}{
  \renewcommand{\arraystretch}{.5}
  \begin{tabular}{@{} crr @{}}
    \toprule
    Design & \# Cells & \# Circuit blocks \\
    \midrule
    vga\_lcd & 56,031 & 15 \\
    OpenPiton & 435,987 & 28 \\
    leon3mp & 374,583 & 50 \\
    netcard & 346,592 & 60 \\
    leon2 & 513,894 & 80 \\
    leon3-avnet & 636,509 & 100 \\
    \bottomrule
  \end{tabular}
  }
  \vspace{-3mm}
  \label{tab:designs}
\end{table}

\begin{table*}[!ht]
\centering
\captionsetup{justification=centering}
\caption{PPAC optimization results in 2 silicon dies (one with 7nm, the other with 45nm). \\
Area in $\times10^3 \mu$m$^2$, WL in $\mu$m, Cost in $\times10^{-3}$, TNS in ns, Power in mW, CPU in seconds, *TSV feasibility}
\vspace{-3mm}
\resizebox{.75\textwidth}{!}{
\renewcommand{\arraystretch}{.5}
\begin{tabular}{@{} lll rrrrrr rrrrrrr @{}}
\toprule
\multirow{2}{*}{Design} & \multicolumn{2}{c}{\multirow{2}{*}{Method}} & \multicolumn{6}{c}{2.5D} & \multicolumn{7}{c}{3D}\\
\cmidrule(lr){4-9} \cmidrule(l){10-16}
& & & Area & WL & Cost & TNS & Power & CPU & Area & WL & Cost & TNS & Power & TSV* & CPU \\
\midrule
\multirow{4}{*}{vga\_lcd} & \multicolumn{2}{l}{Baseline} & 97.21 & 244 & 2205 & -80.16 & 180.5 & 125 & 49.98 & 238 & 2230 & -67.02 & 190.9 & \ding{51} & 186 \\
& \multirow{3}{*}{SMUFTA} & SA & 92.61 & 235 & 2052 & -71.33 & 166.0 & 163 & 48.17 & 219 & 2195 & -51.91 & 175.3 & \ding{51} & 231 \\
& & BO & 90.74 & 231 & 2034 & -72.01 & 165.4 & 305 & 47.65 & 215 & 2137 & -48.05 & 172.3 & \ding{51} & 364 \\
& & RL & 87.39 & 224 & 2017 & -70.12 & 161.3 & 229 & 47.04 & 213 & 2124 & -43.76 & 166.5 & \ding{51} & 287 \\
\midrule
\multirow{4}{*}{OpenPiton} & \multicolumn{2}{l}{Baseline} & 391.11 & 6138 & 2682 & -359.02 & 603.5 & 420 & 200.43 & 5816 & 2759 & -310.55 & 630.1 & \ding{51} & 539 \\
& \multirow{3}{*}{SMUFTA} & SA & 372.78 & 5907 & 2633 & -335.90 & 575.7 & 538 & 191.85 & 5637 & 2668 & -287.49 & 604.2 & \ding{51} & 682 \\
& & BO & 368.46 & 5751 & 2617 & -331.42 & 562.0 & 683 & 189.23 & 5580 & 2645 & -282.13 & 602.0 & \ding{51} & 803 \\
& & RL & 355.81 & 5620 & 2602 & -326.71 & 539.6 & 601 & 185.64 & 5492 & 2630 & -278.90 & 589.6 & \ding{51} & 725 \\
\midrule
\multirow{4}{*}{leon3mp} & \multicolumn{2}{l}{Baseline} & 212.64 & 4270 & 2563 & -290.52 & 641.4 & 872 & 108.50 & 4015 & 2595 & -267.34 & 679.3 & \ding{51} & 976 \\
& \multirow{3}{*}{SMUFTA} & SA & 204.72 & 3917 & 2494 & -285.31 & 617.9 & 1046 & 101.92 & 3874 & 2462 & -253.09 & 625.6 & \ding{51} & 1208 \\
& & BO & 201.36 & 3894 & 2480 & -274.57 & 593.4 & 1306 & 98.07 & 3819 & 2437 & -249.60 & 623.3 & \ding{51} & 1497 \\
& & RL & 195.28 & 3756 & 2414 & -268.20 & 580.5 & 1197 & 94.18 & 3751 & 2351 & -240.18 & 619.1 & \ding{51} & 1324 \\
\midrule
\multirow{4}{*}{netcard} & \multicolumn{2}{l}{Baseline} & 220.59 & 5093 & 2607 & -318.19 & 626.1 & 915 & 124.63 & 4703 & 2650 & -298.91 & 649.0 & \ding{51} & 1105 \\
& \multirow{3}{*}{SMUFTA} & SA & 207.63 & 4601 & 2501 & -291.02 & 609.5 & 1185 & 120.49 & 4512 & 2617 & -276.43 & 625.2 & \ding{51} & 1379 \\
& & BO & 213.01 & 4607 & 2536 & -305.84 & 617.0 & 1391 & 122.87 & 4603 & 2631 & -281.71 & 639.5 & \ding{51} & 1506 \\
& & RL & 192.15 & 4560 & 2395 & -260.13 & 586.4 & 1260 & 113.56 & 4439 & 2583 & -265.02 & 613.7 & \ding{51} & 1482 \\
\midrule
\multirow{4}{*}{leon2} & \multicolumn{2}{l}{Baseline} & 530.92 & 9813 & 2771 & -503.96 & 825.7 & 1408 & 289.15 & 9405 & 2904 & -410.54 & 842.8 & \ding{51} & 1731 \\
& \multirow{3}{*}{SMUFTA} & SA & 505.76 & 9472 & 2609 & -472.52 & 793.9 & 1857 & 272.90 & 9197 & 2811 & -400.13 & 816.9 & \ding{51} & 2185 \\
& & BO & 493.12 & 9386 & 2571 & -460.30 & 758.6 & 2380 & 268.35 & 8954 & 2693 & -395.42 & 803.6 & \ding{51} & 2507 \\
& & RL & 491.08 & 9304 & 2532 & -441.91 & 764.5 & 1743 & 263.82 & 8863 & 2647 & -388.71 & 792.0 & \ding{51} & 1894 \\
\midrule
\multirow{4}{*}{leon3-avnet} & \multicolumn{2}{l}{Baseline} & 891.21 & 11751 & 2964 & -780.65 & 1180.4 & 1831 & 480.17 & 11385 & 2983 & -725.06 & 1207.1 & \ding{51} & 2496 \\
& \multirow{3}{*}{SMUFTA} & SA & 817.74 & 11265 & 2841 & -745.03 & 1105.8 & 2152 & 454.03 & 10921 & 2940 & -683.19 & 1153.6 & \ding{51} & 2913 \\
& & BO & 814.60 & 11203 & 2823 & -736.82 & 1092.5 & 3209 & 447.90 & 10763 & 2934 & -675.50 & 1147.2 & \ding{51} & 3504 \\
& & RL & 803.15 & 11137 & 2788 & -701.49 & 1064.3 & 1987 & 436.75 & 10352 & 2915 & -664.28 & 1128.9 & \ding{51} & 2677 \\
\midrule \midrule
\multirow{4}{*}{\begin{tabular}{@{}c@{}}Norm. \\ Avg.\end{tabular}} & \multicolumn{2}{l}{Baseline} & 1 & 1 & 1 & 1 & 1 & 1 & 1 & 1 & 1 & 1 & 1 & & 1$\times$ \\
& \multirow{3}{*}{SMUFTA} & SA & 0.947 & 0.945 & 0.957 & 0.936 & 0.951 & \textbf{1.26$\times$} & 0.953 & 0.959 & 0.974 & 0.915 & 0.948 & & \textbf{1.24$\times$} \\
& & BO & 0.939 & 0.934 & 0.952 & 0.929 & 0.932 & 1.75$\times$ & 0.940 & 0.947 & 0.960 & 0.899 & 0.941 & & 1.53$\times$ \\
& & RL & \textbf{0.904} & \textbf{0.918} & \textbf{0.933} & \textbf{0.884} & \textbf{0.910} & 1.39$\times$ & \textbf{0.911} & \textbf{0.928} & \textbf{0.946} & \textbf{0.867} & \textbf{0.923} & & 1.29$\times$ \\
\bottomrule
\end{tabular}
}
\vspace{-5mm}
\label{tab:2die_optimization}
\end{table*}

The testcases \change{consist} synthesizable HDL code for five designs in IWLS 2005 benchmarks~\cite{albrecht2005iwls} and a RISC-V-based OpenPiton multi-core system~\cite{balkind2016Openpiton} configured as a $2\times2$ processor using default parameters.
Each design is divided into circuit blocks based on hierarchy (using Synopsys Design Compiler).
The number of cells and blocks are summarized in Table~\ref{tab:designs}.
Experiments \change{utilize} two public-domain technologies, 45nm~\cite{Stine2007freepdk45} and 7nm~\cite{Vashishtha2017asap7}.
\change{Unlike common floorplanning benchmarks, which often lack the associated HDL code, our testcases provide the necessary input for both SMUFTA and the PPA models.}
We compare the following \change{four} floorplanning techniques:
\vspace{-2mm}
\begin{itemize}
\item \textbf{Baseline.} 
\change{Since there is no prior work on 2.5D/3D floorplanning with technology assignment, to the best of our knowledge,
we proposed a baseline that employs greedy partitioning and assignment.
It uses hMetis~\cite{harypis1997HMetis} to partition the interconnected blocks $\mathcal{B}$ into $|\mathcal{D}|$ disjoint subsets minimizing cut sizes.
Next, a greedy step assigns each subset to a die with a technology $g_i \in g_1, \dots, g_k$ that minimizes the partial objective $f_{pt}$ in Eq.~\eqref{eq:part_objective}. The die assignment is completed without replacement.
Then, SA-based floorplanning (Section~\ref{sec:sa}) is performed on each die.
Even though this SA-based baseline lacks inter-die refinement or subsequent re-assignment, it establishes a simple and feasible solution despite its sub-optimality.}

\item \textbf{SMUFTA-SA} is our SMUFTA method that employs SA for intra-die floorplanning. The initial temperature is set to $400$, the cooling factor to $0.85$, and a convergence stopping criteria to $10^{-4}$.

\item \textbf{SMUFTA-BO} employs BO for intra-die floorplanning. The initial evaluations is set to $80$, surrogate model is a Gaussian Process with Mat{\'e}rn kernel, acquisition function is Expected Improvement with a jitter of $10^{-2}$, and the stopping criteria is set to $10^{-4}$.

\item \textbf{SMUFTA-RL} uses RL for intra-die floorplanning. In the PPO, the hyperparameter $\lambda$ is set to $0.2$, the discount factor $\eta$ to $0.95$, the feature selection height $h$ to $6$, and stopping criteria to $10^{-4}$.

\end{itemize}

The objective function weights $(\omega, \beta, \gamma, \tau)$ are set to $(1, 1, 0.5, 2)$.
The bound for inter-die nets $N_{max}$ is $30$.
The bounds $A_{min}$ and $A_{max}$ are set to $0.8z$ and $1.2z$, respectively, where $z$ is the average die area.
The frequency of refinements $K$ is set to $20$.
The total number of optimization steps (moves/evaluations/actions plus inter-die refinements) is limited to $2,400$.
Note that the techniques are adapted to perform 2.5D and 3D floorplanning using the formulations in Section~\ref{sec:formulation}.

The SMUFTA methods are implemented in Python language.
The PPO uses the Spinning Up~\cite{achiam2018spinningup} and Gymnasium~\cite{towers2024gymnasium} libraries. 
Once an optimized SMUFTA solution is obtained, the block floorplan, die assignment, and aspect ratios are used to perform synthesis of each HDL block in Synopsys Design Compiler.
Then, floorplanning and place-and-route is performed for each die in Cadence Innovus.
The heterogeneous integration (interposer in 2.5D/face-to-back in 3D) is completed in Cadence Integrity.
Area, timing, and wirelength are reported at post-routing analysis, power is the summation of each block's power, and cost is calculated using the model in Section~\ref{sec:cost_model}.
Runtime accounts only for our method. The \change{experiments} were performed on an Intel Core i7-1065 CPU with 16GB RAM.

\vspace{-3mm}
\subsection{Results on PPAC Optimization with 2 Dies}

\begin{table*}[!ht]
\centering
\captionsetup{justification=centering}
\caption{PPAC optimization results for leon3-avnet design in 4 silicon dies.\\
Area in $\times10^3 \mu$m$^2$, WL in $\mu$m, Cost in $\times10^{-3}$, TNS in ns, Power in mW, CPU in mins, *TSV feasibility}
\vspace{-3mm}
\resizebox{.75\textwidth}{!}{
\renewcommand{\arraystretch}{0.5}
\begin{tabular}{@{} >{\centering\arraybackslash}p{1.1em}>{\centering\arraybackslash}p{1.1em}ll rrrrrr rrrrrrr @{}}
\toprule
\multicolumn{2}{c}{Dies} & \multicolumn{2}{c}{\multirow{2}{*}{Method}} & \multicolumn{6}{c}{2.5D} & \multicolumn{7}{c}{3D} \\
\cmidrule(lr){1-2} \cmidrule(lr){5-10} \cmidrule(l){11-17}
7nm & 45nm & & & Area & WL & Cost & TNS & Power & CPU & Area & WL & Cost & TNS & Power & TSV* & CPU \\
\midrule

\multirow{4}{*}{1} & \multirow{4}{*}{3} & \multicolumn{2}{l}{Baseline} & 1513.69 & 17355 & 3286 & -1150.81 & 1525.2 & 36.5 & 390.58 & 16953 & 3381 & -1013.57 & 1581.6 & \ding{51} & 39.9 \\
& & \multirow{3}{*}{SMUFTA} & SA & 1433.96 & 16704 & 3152 & -1047.04 & 1447.1 & 40.4 & 372.93 & 16035 & 3209 & -926.21 & 1481.9 & \ding{51} & 40.8 \\
& & & BO & 1458.71 & 16816 & 3180 & -1065.90 & 1460.3 & 55.0 & 374.16 & 16140 & 3255 & -937.04 & 1507.1 & \ding{51} & 56.7 \\
& & & RL & 1359.09 & 16083 & 3092 & -1023.77 & 1395.6 & 33.9 & 353.03 & 15602 & 3147 & -892.50 & 1430.6 & \ding{51} & 36.5 \\
\midrule

\multirow{4}{*}{2} & \multirow{4}{*}{2} & \multicolumn{2}{l}{Baseline} & 910.48 & 13206 & 3026 & -794.59 & 1217.8 & 40.2 & 242.50 & 12825 & 3205 & -720.93 & 1305.3 & \ding{51} & 44.0 \\
& & \multirow{3}{*}{SMUFTA} & SA & 853.51 & 12837 & 2905 & -740.05 & 1180.7 & 44.9 & 231.82 & 12307 & 3028 & -654.39 & 1220.9 & \ding{51} & 47.5 \\
& & & BO & 867.43 & 12910 & 2941 & -746.78 & 1163.2 & 59.2 & 235.17 & 12415 & 3054 & -677.42 & 1263.4 & \ding{51} & 61.2 \\
& & & RL & 829.02 & 12404 & 2798 & -723.95 & 1104.0 & 38.1 & 224.69 & 11893 & 2962 & -636.74 & 1186.0 & \ding{51} & 39.2 \\
\midrule

\multirow{4}{*}{3} & \multirow{4}{*}{1} & \multicolumn{2}{l}{Baseline} & 627.20 & 9602 & 2729 & -640.16 & 702.3 & 44.6 & 165.40 & 9492 & 2924 & -568.52 & 764.0 & \ding{51} & 46.8 \\
& & \multirow{3}{*}{SMUFTA} & SA & 578.24 & 9216 & 2561 & -591.02 & 673.0 & 47.4 & 161.03 & 9144 & 2801 & -516.31 & 728.0 & \ding{51} & 49.3 \\
& & & BO & 592.50 & 9307 & 2609 & -612.74 & 685.1 & 62.2 & 163.91 & 9270 & 2863 & -530.71 & 741.5 & \ding{51} & 61.4 \\
& & & RL & 561.39 & 8944 & 2513 & -571.30 & 649.3 & 41.8 & 157.30 & 8901 & 2742 & -497.28 & 691.7 & \ding{51} & 42.2 \\
\midrule \midrule

& & \multicolumn{2}{l}{Baseline} & 1 & 1 & 1 & 1 & 1 & 1 & 1 & 1 & 1 & 1 & 1 & & 1$\times$ \\
\multicolumn{2}{l}{Norm.} & \multirow{3}{*}{SMUFTA} & SA & 0.936 & 0.965 & 0.953 & 0.921 & 0.959 & 1.10$\times$ & 0.961 & 0.956 & 0.951 & 0.910 & 0.942 & & 1.05$\times$ \\
\multicolumn{2}{l}{Avg.} & & BO & 0.955 & 0.972 & 0.965 & 0.941 & 0.964 & 1.46$\times$ & 0.975 & 0.968 & 0.964 & 0.933 & 0.965 & & 1.37$\times$ \\
& & & RL & \textbf{0.901} & \textbf{0.932} & \textbf{0.929} & \textbf{0.898} & \textbf{0.915} & \textbf{0.94$\times$} & \textbf{0.927} & \textbf{0.928} & \textbf{0.931} & \textbf{0.879} & \textbf{0.906} & & \textbf{0.90$\times$} \\
\bottomrule
\end{tabular}%
}
\vspace{-3mm}
\label{tab:4die_optimization}
\end{table*}


Table~\ref{tab:2die_optimization} summarizes the results for the four methods using 2 dies (one with 7nm, and the other with 45nm).
In the chiplet integration, SMUFTA-RL achieves the best overall results, with average reductions of 11.6\% in performance, 9\% in power, 9.6\% in area, 6.7\% in cost, and 8.2\% in post-routing wirelength compared to the baseline.

In the 3D IC configuration, SMUFTA-RL again performs better than others, achieving average reductions of 13.3\% in performance, 7.7\% in power, 8.9\% in area, 5.4\% in cost, and 7.2\% in wirelength, while ensuring TSV feasibility.
SMUFTA-RL consistently provides better optimization quality in both 2.5D and 3D formulations with a moderate \change{runtime} overhead.

\vspace{-3mm}
\subsection{Results on PPAC Optimization with 4 Dies}

\begin{table*}[!ht]
\centering
\captionsetup{justification=centering}
\caption{PPAC optimization results with hard IPs for netcard design in 2 silicon dies (one with 7nm, the other with 45nm).\\
Area in $\times10^3 \mu$m$^2$, WL in $\mu$m, Cost in $\times10^{-3}$, TNS in ns, Power in mW, *TSV feasibility}
\vspace{-3mm}
\resizebox{0.6\textwidth}{!}{
\renewcommand{\arraystretch}{0.5}
\begin{tabular}{@{}>{\centering\arraybackslash}p{1.1em}ll rrrrr rrrrrr@{}}
\toprule
\multirow{2}{*}{\begin{tabular}{@{}c@{}}Hard \\ IPs\end{tabular}} & \multicolumn{2}{c}{\multirow{2}{*}{Method}} & \multicolumn{5}{c}{2.5D} & \multicolumn{6}{c}{3D} \\
\cmidrule(lr){4-8} \cmidrule(l){9-14}
& & & Area & WL & Cost & TNS & Power & Area & WL & Cost & TNS & Power & TSV* \\
\midrule
\multirow{2}{*}{5} & \multicolumn{2}{l}{Baseline} & 227.52 & 5228 & 2715 & -341.68 & 671.0 & 130.95 & 5394 & 2903 & -314.92 & 690.2 & \ding{51} \\
& SMUFTA & RL & 203.30 & 4816 & 2495 & -301.42 & 613.2 & 122.17 & 5082 & 2797 & -275.03 & 652.1 & \ding{51} \\
\midrule
\multirow{2}{*}{10} & \multicolumn{2}{l}{Baseline} & 235.68 & 5451 & 2763 & -355.29 & 701.9 & 138.62 & 5531 & 3008 & -320.61 & 716.6 & \ding{51} \\
& SMUFTA & RL & 207.95 & 4910 & 2517 & -309.47 & 636.0 & 129.58 & 5127 & 2861 & -287.98 & 683.5 & \ding{51} \\
\midrule
\multirow{2}{*}{15} & \multicolumn{2}{l}{Baseline} & 246.02 & 5594 & 2811 & -390.31 & 730.3 & 147.02 & 5628 & 3055 & -361.74 & 741.9 & \ding{51} \\
& SMUFTA & RL & 218.43 & 5017 & 2590 & -337.08 & 652.8 & 136.31 & 5169 & 2943 & -337.19 & 695.0 & \ding{51} \\
\midrule \midrule
\multicolumn{3}{c}{Norm. Avg.} & 0.887 & 0.906 & 0.918 & 0.872 & 0.903 & 0.932 & 0.929 & 0.957 & 0.901 & 0.947 & \\
\bottomrule
\end{tabular}
}
\vspace{-4mm}
\label{tab:fixed_ips}
\end{table*}


Table~\ref{tab:4die_optimization} shows the results for the \mbox{\emph{leon3-avnet}} design using 4 dies. In the 2.5D integration, SMUFTA-RL shows the best average results, achieving reductions of 10.2\% in performance, 8.5\% in power, 9.9\% in area, 7.1\% in cost, and 6.8\% in post-routing wirelength compared to the baseline.
Also, SMUFTA-RL runtime is $0.94\times$ faster than the \change{runtime of} baseline, whereas SA ($1.10\times$) and BO ($1.46\times$) are slower.

In the 3D IC configuration, SMUFTA-RL again stands out, showing reductions of 12.1\% in performance, 9.4\% in power, 7.3\% in area, 6.9\% in cost, and 7.2\% in wirelength, also ensuring TSV feasibility.

\vspace{-3mm}
\subsection{SMUFTA-RL with Hard IPs}

Table~\ref{tab:fixed_ips} shows results for the \mbox{\emph{netcard}} design using 2 dies where some blocks are set as hard IPs in the 45nm node (fixed technology and aspect ratio). In the chiplet formulation, SMUFTA-RL achieves the best average results showing improvements of 12.8\% in performance, 9.7\% in power, 11.3\% in area, 8.2\% in cost, and 9.4\% in post-routing wirelength compared to the baseline.
For the 3D formulation, SMUFTA-RL also shows improvements over the other techniques, and maintains TSV feasibility even with hard IPs.

\vspace{-2mm}
\subsection{Timing-Power Tradeoff}
\begin{figure}[!ht] 
\centering
\vspace{-5mm}
\includegraphics[width=.8\linewidth]{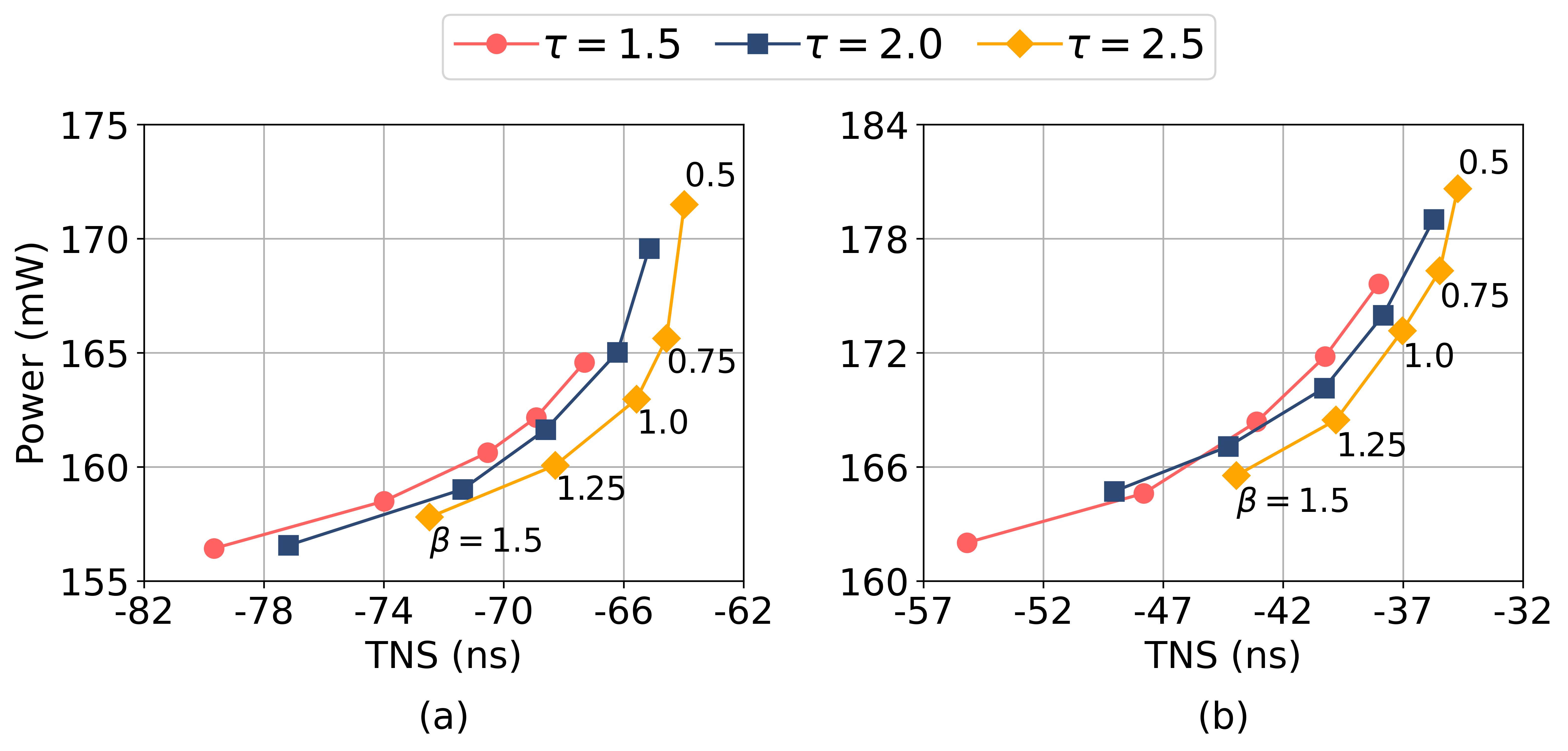}
\vspace{-4mm}
\caption{TNS-power tradeoff on the \emph{vga\_lcd}, varying timing ($\tau$) and power ($\beta$) weights in (a) chiplet and (b) 3D configurations.}
\vspace{-5mm}
\label{fig:power_timing_tradeoff}
\end{figure}

Figure~\ref{fig:power_timing_tradeoff} shows the TNS-power tradeoff for SMUFTA-RL on the \emph{vga\_lcd} design.
The objective $f$ uses the weights $\tau$ (TNS) and $\beta$ (power).
As $\beta$ increases, power consumption decreases while TNS worsens.
Likewise, as $\tau$ increases, TNS improves (less negative) while power increases. Thus, the weighting factors can be tuned to achieve a different timing-power tradeoff.

\vspace{-2mm}
\subsection{Impact of Inter-die Refinement Frequency \texorpdfstring{$K$}{K}}
\begin{figure}[!ht] 
\centering
\includegraphics[width=.8\linewidth]{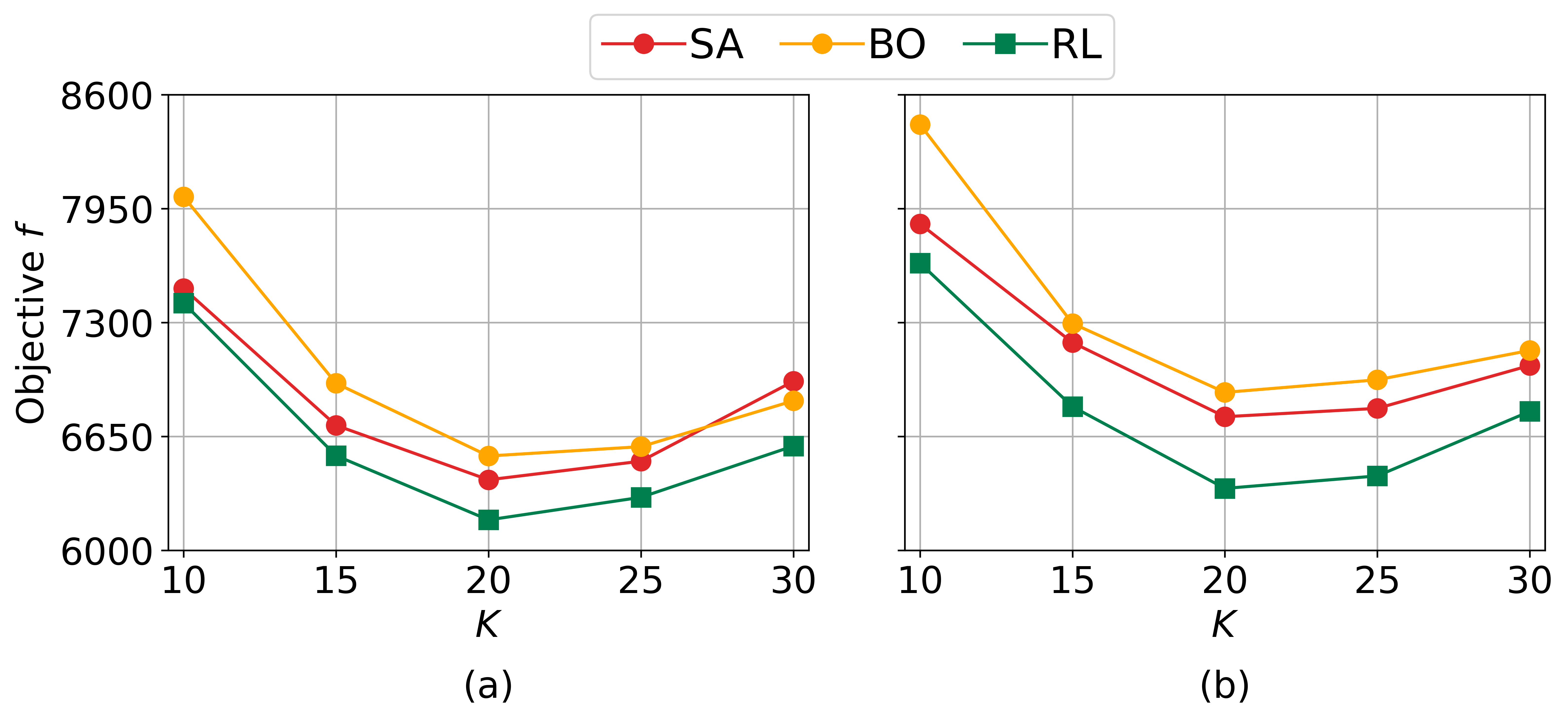}
\vspace{-3mm}
\caption{Objective $f$ vs. refinement frequency $K$ for SMUFTA on the \emph{netcard} in (a) 2.5D and (b) 3D configurations.}
\vspace{-3mm}
\label{fig:step_K}
\end{figure}

Figure~\ref{fig:step_K} shows the objective $f$ versus refinement frequency $K$ for SMUFTA-SA/BO/RL on the \emph{netcard} design.
SMUFTA is performed for both chiplet and 3D formulations (Figure~\ref{fig:step_K}a~and~b, respectively).
The objective degrades when $K$ is either too small or too large, and obtaining best results when $K=20$.
The value of $K$ impacts CPU runtime: small value causes frequent interruptions in intra-die floorplanning, while large value prevents convergence before the iteration limit.

\vspace{-2mm}
\subsection{Importance of Using an RL Agent per Die}
\begin{figure}[!ht] 
\centering
\includegraphics[width=.85\linewidth]{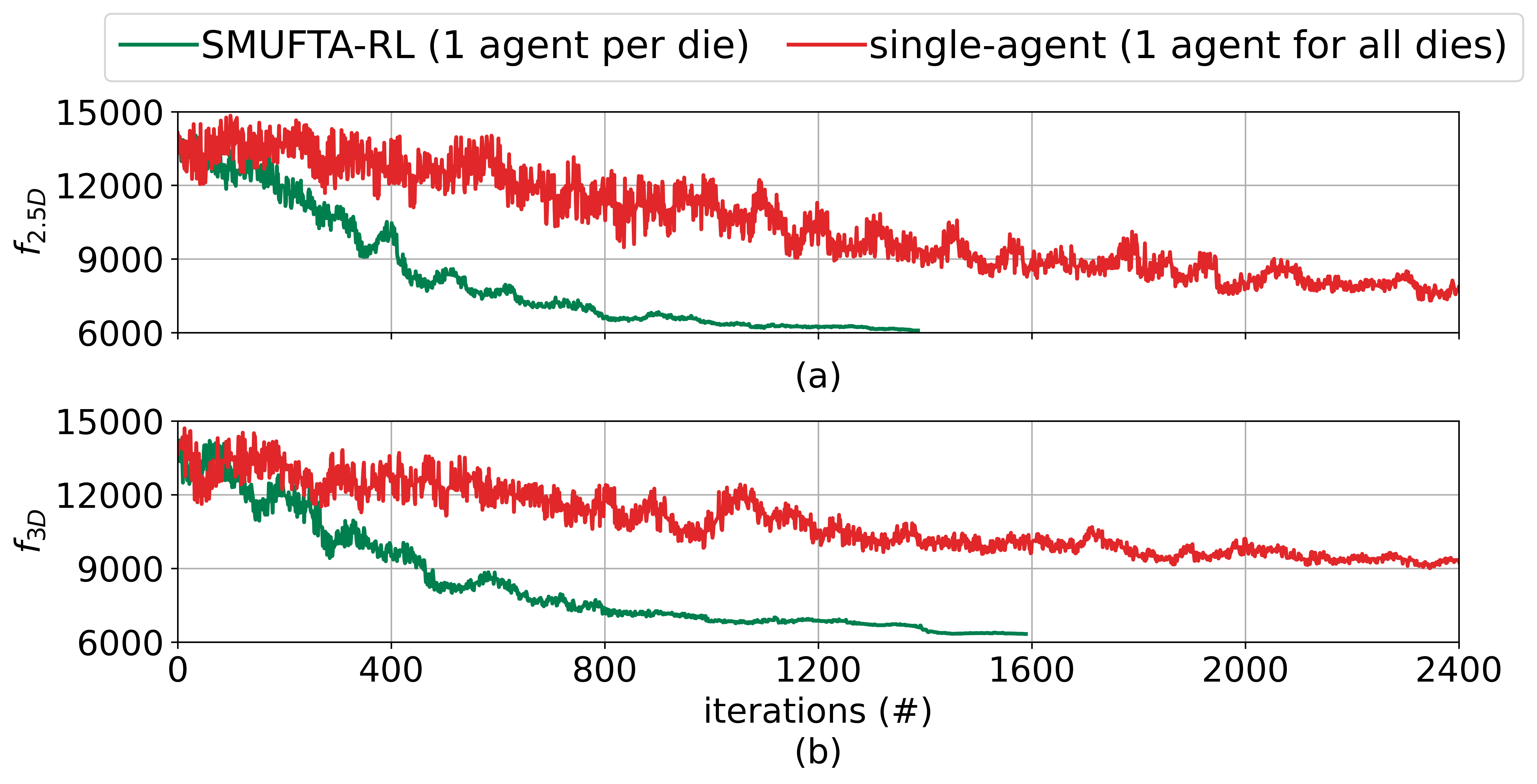}
\vspace{-3mm}
\caption{Objective $f$ convergence comparing SMUFTA-RL vs single-agent on \emph{netcard} in (a) chiplet and (b) 3D formulations.}
\vspace{-3mm}
\label{fig:agent_iteration}
\end{figure}

SMUFTA-RL employs a dedicated RL agent per die (Section \ref{sec:rl}).
We evaluate our SMUFTA-RL (1 agent per die) and a single-agent approach (1 agent for all dies) on the \emph{netcard} design, the single-agent fails to converge and shows high variability (Figure~\ref{fig:agent_iteration}).
This result emphasizes the importance of using a dedicated agent for each die to leverage die-specific technology information.

\vspace{-2mm}
\section{Conclusions}
\label{sec:conclusions}
This work provides the first study on simultaneous multi-die floorplanning and technology assignment for 2.5D and 3D IC designs, to the best of our knowledge, as almost no prior work considered simultaneous technology assignment.
Post-routing analysis using commercial tools demonstrates that the proposed techniques outperforms a greedy baseline in terms of wirelength, area, power, timing and die cost.
In our future research, we plan to extend this method to additionally address thermal and warpage issues, \change{and support additional public-domain technology nodes}.

\balance
\bibliographystyle{unsrt}
\bibliography{./references.bib}

@inproceedings{ho2013btreeSA,
author = {Ho, Yuan-Kai and Chang, Yao-Wen},
title = {Multiple chip planning for chip-interposer codesign},
year = {2013},
isbn = {9781450320719},
publisher = {Association for Computing Machinery},
address = {New York, NY, USA},
url = {https://doi.org/10.1145/2463209.2488767},
doi = {10.1145/2463209.2488767},
booktitle = {Proceedings of the 50th Annual Design Automation Conference},
articleno = {27},
numpages = {6},
location = {Austin, Texas},
series = {DAC '13}}

@inproceedings{lee2023multiDieFloorplan,
  author={Lee, Chung-Chia and Chang, Yao-Wen},
  booktitle={2023 IEEE/ACM International Conference on Computer Aided Design (ICCAD)}, 
  title={Floorplanning for Embedded Multi-Die Interconnect Bridge Packages}, 
  year={2023},
  volume={},
  number={},
  pages={1-8},
  doi={10.1109/ICCAD57390.2023.10323609}}

@inproceedings{sengupta2022ppa,
author = {Sengupta, Prianka and Tyagi, Aakash and Chen, Yiran and Hu, Jiang},
title = {How Good Is Your Verilog RTL Code? A Quick Answer from Machine Learning},
year = {2022},
isbn = {9781450392174},
publisher = {Association for Computing Machinery},
address = {New York, NY, USA},
url = {https://doi.org/10.1145/3508352.3549375},
doi = {10.1145/3508352.3549375},
booktitle = {Proceedings of the 41st IEEE/ACM International Conference on Computer-Aided Design},
articleno = {89},
numpages = {9},
location = {San Diego, California},
series = {ICCAD '22}}

@ARTICLE{chen2024floorplet,
  author={Chen, Shixin and Li, Shanyi and Zhuang, Zhen and Zheng, Su and Liang, Zheng and Ho, Tsung-Yi and Yu, Bei and Sangiovanni-Vincentelli, Alberto L.},
  journal={IEEE Transactions on Computer-Aided Design of Integrated Circuits and Systems}, 
  title={Floorplet: Performance-Aware Floorplan Framework for Chiplet Integration}, 
  year={2024},
  volume={43},
  number={6},
  pages={1638-1649},
  doi={10.1109/TCAD.2023.3347302}}

@inproceedings{hsu2022warpageFloorplan,
author = {Hsu, Yang and Chung, Min-Hsuan and Chang, Yao-Wen and Lin, Ci-Hong},
title = {Transitive Closure Graph-Based Warpage-Aware Floorplanning for Package Designs},
year = {2022},
isbn = {9781450392174},
publisher = {Association for Computing Machinery},
address = {New York, NY, USA},
url = {https://doi.org/10.1145/3508352.3549354},
doi = {10.1145/3508352.3549354},
booktitle = {Proceedings of the 41st IEEE/ACM International Conference on Computer-Aided Design},
articleno = {16},
numpages = {7},
location = {San Diego, California},
series = {ICCAD '22}}

@ARTICLE{xu2022goodfloorplan,
  author={Xu, Qi and Geng, Hao and Chen, Song and Yuan, Bo and Zhuo, Cheng and Kang, Yi and Wen, Xiaoqing},
  journal={IEEE Transactions on Computer-Aided Design of Integrated Circuits and Systems}, 
  title={GoodFloorplan: Graph Convolutional Network and Reinforcement Learning-Based Floorplanning}, 
  year={2022},
  volume={41},
  number={10},
  pages={3492-3502},
  doi={10.1109/TCAD.2021.3131550}}

@INPROCEEDINGS{ma2021thermalFloorplan,
  author={Ma, Yenai and Delshadtehrani, Leila and Demirkiran, Cansu and Abellan, José L. and Joshi, Aiav},
  booktitle={2021 Design, Automation \& Test in Europe Conference \& Exhibition (DATE)}, 
  title={TAP-2.5D: A Thermally-Aware Chiplet Placement Methodology for 2.5D Systems}, 
  year={2021},
  volume={},
  number={},
  pages={1246-1251},
  doi={10.23919/DATE51398.2021.9474011}}

@INPROCEEDINGS{duan2024RLPlanner,
  author={Duan, Yuanyuan and Liu, Xingchen and Yu, Zhiping and Wu, Hanming and Shao, Leilai and Zhu, Xiaolei},
  booktitle={2024 Design, Automation \& Test in Europe Conference \& Exhibition (DATE)}, 
  title={RLPlanner: Reinforcement Learning Based Floorplanning for Chiplets with Fast Thermal Analysis}, 
  year={2024},
  volume={},
  number={},
  pages={1-2},
  doi={10.23919/DATE58400.2024.10546812}}

@misc{amin2024RLFloorplan3D,
  title={Large Reasoning Models for 3D Floorplanning in EDA: Learning from Imperfections}, 
  author={Fin Amin and Nirjhor Rouf and Tse-Han Pan and Md Kamal Ibn Shafi and Paul D. Franzon},
  year={2024},
  eprint={2406.10538},
  archivePrefix={arXiv},
  primaryClass={cs.LG},
  url={https://arxiv.org/abs/2406.10538}, }

@inproceedings{chang2000btree,
author = {Chang, Yun-Chih and Chang, Yao-Wen and Wu, Guang-Ming and Wu, Shu-Wei},
title = {B*-Trees: a new representation for non-slicing floorplans},
year = {2000},
isbn = {1581131879},
publisher = {Association for Computing Machinery},
address = {New York, NY, USA},
url = {https://doi.org/10.1145/337292.337541},
doi = {10.1145/337292.337541},
booktitle = {Proceedings of the 37th Annual Design Automation Conference},
pages = {458–463},
numpages = {6},
location = {Los Angeles, California, USA},
series = {DAC '00}}

@inproceedings{harypis1997HMetis,
author = {Karypis, George and Aggarwal, Rajat and Kumar, Vipin and Shekhar, Shashi},
title = {Multilevel hypergraph partitioning: application in VLSI domain},
year = {1997},
isbn = {0897919203},
publisher = {Association for Computing Machinery},
address = {New York, NY, USA},
url = {https://doi.org/10.1145/266021.266273},
doi = {10.1145/266021.266273},
booktitle = {Proceedings of the 34th Annual Design Automation Conference},
pages = {526–529},
numpages = {4},
location = {Anaheim, California, USA},
series = {DAC '97}}

@ARTICLE{cunningham1990costModel,
  author={Cunningham, J.A.},
  journal={IEEE Transactions on Semiconductor Manufacturing}, 
  title={The use and evaluation of yield models in integrated circuit manufacturing}, 
  year={1990},
  volume={3},
  number={2},
  pages={60-71},
  doi={10.1109/66.53188}}

@inproceedings{chang2024multidieChallenges,
author = {Chang, Yao-Wen},
title = {Physical Design Challenges in Modern Heterogeneous Integration},
year = {2024},
isbn = {9798400704178},
publisher = {Association for Computing Machinery},
address = {New York, NY, USA},
url = {https://doi.org/10.1145/3626184.3639690},
doi = {10.1145/3626184.3639690},
booktitle = {Proceedings of the 2024 International Symposium on Physical Design},
pages = {125–134},
numpages = {10},
location = {Taipei, Taiwan},
series = {ISPD '24}}

@inproceedings{zhuang2022multipackage,
author = {Zhuang, Zhen and Yu, Bei and Chao, Kai-Yuan and Ho, Tsung-Yi},
title = {Multi-Package Co-Design for Chiplet Integration},
year = {2022},
isbn = {9781450392174},
publisher = {Association for Computing Machinery},
address = {New York, NY, USA},
url = {https://doi.org/10.1145/3508352.3549404},
doi = {10.1145/3508352.3549404},
booktitle = {Proceedings of the 41st IEEE/ACM International Conference on Computer-Aided Design},
articleno = {114},
numpages = {9},
location = {San Diego, California},
series = {ICCAD '22}}

@inproceedings{feng2022costModel,
author = {Feng, Yinxiao and Ma, Kaisheng},
title = {Chiplet actuary: a quantitative cost model and multi-chiplet architecture exploration},
year = {2022},
isbn = {9781450391429},
publisher = {Association for Computing Machinery},
address = {New York, NY, USA},
url = {https://doi.org/10.1145/3489517.3530428},
doi = {10.1145/3489517.3530428},
booktitle = {Proceedings of the 59th ACM/IEEE Design Automation Conference},
pages = {121–126},
numpages = {6},
location = {San Francisco, California},
series = {DAC '22}}

@INPROCEEDINGS{osmolovskyi2018diePlacement,
  author={Osmolovskyi, Sergii and Knechtel, Johann and Markov, Igor L. and Lienig, Jens},
  booktitle={2018 23rd Asia and South Pacific Design Automation Conference (ASP-DAC)}, 
  title={Optimal die placement for interposer-based 3D ICs}, 
  year={2018},
  volume={},
  number={},
  pages={513-520},
  doi={10.1109/ASPDAC.2018.8297375}}

@misc{schulman2017PPOclip,
  title={Proximal Policy Optimization Algorithms}, 
  author={John Schulman and Filip Wolski and Prafulla Dhariwal and Alec Radford and Oleg Klimov},
  year={2017},
  eprint={1707.06347},
  archivePrefix={arXiv},
  primaryClass={cs.LG},
  url={https://arxiv.org/abs/1707.06347}, }

@inproceedings{albrecht2005iwls,
  title={IWLS 2005 benchmarks},
  author={{Albrecht, Christoph}},
  booktitle={International Workshop for Logic Synthesis (IWLS)},
  volume={9},
  year={2005}}

@inproceedings{balkind2016Openpiton,
author = {Balkind, Jonathan and McKeown, Michael and Fu, Yaosheng and Nguyen, Tri and Zhou, Yanqi and Lavrov, Alexey and Shahrad, Mohammad and Fuchs, Adi and Payne, Samuel and Liang, Xiaohua and Matl, Matthew and Wentzlaff, David},
title = {OpenPiton: An Open Source Manycore Research Framework},
year = {2016},
isbn = {9781450340915},
publisher = {Association for Computing Machinery},
address = {New York, NY, USA},
url = {https://doi.org/10.1145/2872362.2872414},
doi = {10.1145/2872362.2872414},
booktitle = {Proceedings of the Twenty-First International Conference on Architectural Support for Programming Languages and Operating Systems},
pages = {217–232},
numpages = {16},
location = {Atlanta, Georgia, USA},
series = {ASPLOS '16}}

@misc{achiam2018spinningup,
  author = {Achiam, Joshua},
  title = {Spinning Up in Deep Reinforcement Learning},
  year = {2018}}

@article{towers2024gymnasium,
  title={Gymnasium: A Standard Interface for Reinforcement Learning Environments},
  author={Towers, Mark and Kwiatkowski, Ariel and Terry, Jordan and Balis, John U and De Cola, Gianluca and Deleu, Tristan and Goul{\~a}o, Manuel and Kallinteris, Andreas and Krimmel, Markus and KG, Arjun and others},
  journal={arXiv preprint arXiv:2407.17032},
  year={2024}}

@INPROCEEDINGS{Vashishtha2017asap7,
  author={Vashishtha, Vinay and Vangala, Manoj and Clark, Lawrence T.},
  booktitle={2017 IEEE/ACM International Conference on Computer-Aided Design (ICCAD)}, 
  title={ASAP7 predictive design kit development and cell design technology co-optimization: Invited paper}, 
  year={2017},
  volume={},
  number={},
  pages={992-998},
  doi={10.1109/ICCAD.2017.8203889}}

@INPROCEEDINGS{Stine2007freepdk45,
  author={Stine, James E. and Castellanos, Ivan and Wood, Michael and Henson, Jeff and Love, Fred and Davis, W. Rhett and Franzon, Paul D. and Bucher, Michael and Basavarajaiah, Sunil and Oh, Julie and Jenkal, Ravi},
  booktitle={2007 IEEE International Conference on Microelectronic Systems Education (MSE'07)}, 
  title={FreePDK: An Open-Source Variation-Aware Design Kit}, 
  year={2007},
  volume={},
  number={},
  pages={173-174},
  doi={10.1109/MSE.2007.44}}

@INPROCEEDINGS{wong1986polishExpression,
  author={Wong, D.F. and Liu, C.L.},
  booktitle={23rd ACM/IEEE Design Automation Conference}, 
  title={A New Algorithm for Floorplan Design}, 
  year={1986},
  volume={},
  number={},
  pages={101-107},
  doi={10.1109/DAC.1986.1586075}}

@Inbook{murata2003sequencePair,
author="Murata, Hiroshi
and Fujiyoshi, Kunihiro
and Nakatake, Shigetoshi
and Kajitani, Yoji",
editor="Kuehlmann, Andreas",
title="Rectangle-Packing-Based Module Placement",
bookTitle="The Best of ICCAD: 20 Years of Excellence in Computer-Aided Design",
year="2003",
publisher="Springer US",
address="Boston, MA",
pages="535--548",
isbn="978-1-4615-0292-0",
doi="10.1007/978-1-4615-0292-0\_42",
url="https://doi.org/10.1007/978-1-4615-0292-0\_42"}

@INPROCEEDINGS{nakatake1996boundSlicing,
  author={Nakatake, S. and Fujiyoshi, K. and Murata, H. and Kajitani, Y.},
  booktitle={Proceedings of International Conference on Computer Aided Design}, 
  title={Module placement on BSG-structure and IC layout applications}, 
  year={1996},
  volume={},
  number={},
  pages={484-491},
  doi={10.1109/ICCAD.1996.569870}}

@inproceedings{guo1999oTree,
author = {Guo, Pei-Ning and Cheng, Chung-Kuan and Yoshimura, Takeshi},
title = {An O-tree representation of non-slicing floorplan and its applications},
year = {1999},
isbn = {1581131097},
publisher = {Association for Computing Machinery},
address = {New York, NY, USA},
url = {https://doi.org/10.1145/309847.309928},
doi = {10.1145/309847.309928},
booktitle = {Proceedings of the 36th Annual ACM/IEEE Design Automation Conference},
pages = {268–273},
numpages = {6},
location = {New Orleans, Louisiana, USA},
series = {DAC '99}}

@INPROCEEDINGS{hong2000cornerList,
  author={Xianlong Hong and Gang Huang and Yici Cai and Jiangchun Gu and Sheqin Dong and Chung-Kuan Cheng and Jun Gu},
  booktitle={IEEE/ACM International Conference on Computer Aided Design. ICCAD - 2000. IEEE/ACM Digest of Technical Papers (Cat. No.00CH37140)}, 
  title={Corner block list: an effective and efficient topological representation of non-slicing floorplan}, 
  year={2000},
  volume={},
  number={},
  pages={8-12},
  doi={10.1109/ICCAD.2000.896442}}

@ARTICLE{hwang2011tsvFloorplan,
  author={Tsai, Ming-Chao and Wang, Ting-Chi and Hwang, TingTing},
  journal={IEEE Transactions on Very Large Scale Integration (VLSI) Systems}, 
  title={Through-Silicon Via Planning in 3-D Floorplanning}, 
  year={2011},
  volume={19},
  number={8},
  pages={1448-1457},
  doi={10.1109/TVLSI.2010.2050012}}

@ARTICLE{want2013tsvCoPlace,
  author={Li, Cha-Ru and Mak, Wai-Kei and Wang, Ting-Chi},
  journal={IEEE Transactions on Very Large Scale Integration (VLSI) Systems}, 
  title={Fast Fixed-Outline 3-D IC Floorplanning With TSV Co-Placement}, 
  year={2013},
  volume={21},
  number={3},
  pages={523-532},
  doi={10.1109/TVLSI.2012.2190537}}

@inproceedings{bei2025chipletChallenge,
    author = {Chen, Shixin and Zhang, Hengyuan and Ling, Zichao and Zhai, Jianwang and Yu, Bei},
    title = {The Survey of 2.5D Integrated Architecture: An EDA perspective},
    year = {2025},
    isbn = {9798400706356},
    publisher = {Association for Computing Machinery},
    address = {New York, NY, USA},
    url = {https://doi.org/10.1145/3658617.3703134},
    doi = {10.1145/3658617.3703134},
    booktitle = {Proceedings of the 30th Asia and South Pacific Design Automation Conference},
    pages = {285–293},
    numpages = {9},
    location = {Tokyo, Japan},
    series = {ASPDAC '25}
}
\end{document}